\documentclass[aps,pra,preprint,showpacs]{revtex4}
\newcommand{\PRA}[3]{Phys.~Rev.~A       {\bf #1}, #2 (#3).}

\begin{document}
\preprint{Version 2.0}

\title{Extended Hylleraas three-electron integral}

\author{Krzysztof Pachucki}
\email[]{krp@fuw.edu.pl}

\author{Mariusz Puchalski}
\email[]{mpuchals@fuw.edu.pl}

\affiliation{Institute of Theoretical Physics, Warsaw University,
             Ho\.{z}a 69, 00-681 Warsaw, Poland}


\begin{abstract}
A closed form expression for the three-electron Hylleraas integral
involving the inverse quadratic power of one inter-particle coordinate
is obtained, and recursion relations are derived for positive powers of other coordinates.
This result is suited for high precision calculations of
relativistic effects in lithium and light lithium-like ions.
\end{abstract}

\maketitle
\section{Introduction}
The subject of this work is the extended three-electron Hylleraas integrals involving
$1/r_{ij}^2$ and $1/r_i^2$ terms, which appear in matrix elements of 
relativistic operators in the Breit-Pauli Hamiltonian.
These integrals are defined as
\begin{eqnarray}
f(n_1,n_2,n_3,n_4,n_5,n_6) &=& \int\frac{d^3\,r_1}{4\,\pi}\int\frac{d^3\,r_2}{4\,\pi}
\int\frac{d^3\,r_3}{4\,\pi}\,e^{-w_1\,r_1-w_2\,r_2-w_3\,r_3}\nonumber \\ &&
r_{23}^{n_1-1}\,r_{31}^{n_2-1}\,r_{12}^{n_3-1}\,
r_{1}^{n_4-1}\,r_{2}^{n_5-1}\,r_{3}^{n_6-1} ,
\label{01}
\end{eqnarray}
with one of $n_i$ equal to $-1$, and all other $n_i$ are nonnegative integers.
They have been studied in detail in a series of papers by King
\cite{king_lit1, king_lit2,king_lit3} and by Yan {\em et al.} \cite{yan_lit}. 
Their approach is based on the expansion of $1/r_{ij}^n$
in an infinite  series of some orthogonal polynomials. The resulting multiple summation
is performed with the help of convergence accelerators. 
In some special cases \cite{king_lit1} this expansion can be avoided 
and Eq. (\ref{01}) can be expressed in terms of 
the two-electron Hylleraas integral. Using these methods leading relativistic 
and QED corrections to energies \cite{king_rev, hyl_lit},
isotope-shift \cite{lit_iso}, and  $g$-factors \cite{lit_g}
have been calculated to a high degree of accuracy. The analytic approach 
to Hylleraas integral has been so far much less successful. 
Fromm and Hill in \cite{fh} were able
to obtain an analytic expression for a more general integral with exponents
in $r_i$, as well as in $r_{ij}$. However,
their expression is quite complicated, involves multivalued dilogarithmic
functions, and thus is of limited use.
In parallel Remiddi in \cite{remiddi} obtained a simple analytic expression for the
Hylleraas integral with $n_i=0$. Recently, together with Remiddi we derived
recursion relations \cite{rec} for Hylleraas integrals for arbitrary
large powers of $r_i$ and $r_{ij}$. This result allows
for a convenient calculation of the nonrelativistic wave function of the lithium atom
and light lithium-like ions. 

In this work we present an  analytic approach which allows for a
fast and high precision calculation of extended 3-electron Hylleraas
integrals. This sets the ground for improving the precision
of theoretical energy levels by including higher order relativistic 
corrections \cite{fw}. We obtain here the closed form expression for the master integral
$f(-1,0,0,0,0,0)$  and derive recursion relations for increasing values
of arguments $n_i$ of $f(-1,n_2,n_3,n_4,n_5,n_6)$.
The other integrals, such us $f(n_1,n_2,n_3,-1,n_5,n_6)$,
can be  obtained by using already derived recursion relations for
$f(n_1,n_2,n_3,n_4,n_5,n_6)$ with nonnegative $n_i$
followed by a one dimensional numerical integration 
with respect to the corresponding parameter $w_i$. 

\section{Recursion relations for $r_{23}^{-2}$ integral}
Our derivation is based on integration by parts identities which are 
commonly used for the calculations of multi-loop Feynman diagrams \cite{parts}.
We follow here the former work \cite{rec} and first consider the integral $G$
\begin{eqnarray}
G(m_1,m_2,m_3;m_4,m_5,m_6) &=& \frac{1}{8\,\pi^6}\,\int d^3k_1\int d^3k_2\int d^3k_3\,
(k_1^2+u_1^2)^{-m_1}\,(k_2^2+u_2^2)^{-m_2} \nonumber\\ &&(k_3^2+u_3^2)^{-m_3}\,
(k_{32}^2+w_1^2)^{-m_4}\,(k_{13}^2+w_2^2)^{-m_5}\,(k_{21}^2+w_3^2)^{-m_6}, \label{02}
\end{eqnarray}
which is related to $f$ by: $f(0,0,0,0,0,0,0) = G(1,1,1,1,1,1)|_{u_1=u_2=u_3=0}$.
The following 9 integration by part identities are valid
because the integral of the derivative of a function 
vanishing at infinity vanishes,
\begin{eqnarray}
&&0 \equiv {\rm id}(i,j) = 
\int d^3k_1\int d^3k_2\int d^3k_3\,\frac{\partial}{\partial\,{\vec k_i}}
 \Bigl[ \vec k_j\,(k_1^2+u_1^2)^{-1} 
\nonumber \\ &&  
(k_2^2+u_2^2)^{-1}\,(k_3^2+u_3^2)^{-1}
(k_{32}^2+w_1^2)^{-1}\,(k_{13}^2+w_2^2)^{-1}\,(k_{21}^2+w_3^2)^{-1} 
\Bigr] ,
\label{03}
\end{eqnarray} 
where $i,j=1,2,3$.
The reduction of the scalar products from the numerator leads to the 
identities for the linear combination of the $G$-function.
If any of the arguments is equal to 0, then $G$ becomes a known two-electron
Hylleraas type integral, Eq. (\ref{B6}). 
The explicit form of all 9 identities is presented in Eq. (\ref{A1})
and the whole derivation presented below is performed with the help
of {\sl MATHEMATICA} program for symbolic computation.

In the first step of deriving recursion relations
we take the difference ${\rm id}(3,2)-{\rm id}(2,2)$ 
and use it as an equation for $G(1, 2, 1, 1, 1, 1)$,
\begin{eqnarray}
G(1, 2, 1, 1, 1, 1)\,(u_2^2 - u_3^2 + w_1^2) &=& 
                         G(1, 1, 1, 0, 1, 2) - G(1, 1, 1, 1, 0, 2) 
                \nonumber \\ &&
                        + G(1, 1, 1, 1, 1, 1) - G(1, 2, 0, 1, 1, 1) 
                \nonumber \\ &&
                        + G(1, 2, 1, 0, 1, 1) - 2\,G(1, 1, 1, 2, 1, 1)\,w_1^2 
                \nonumber \\ &&
                        + G(1, 1, 1, 1, 1, 2)\,(w_2^2 - w_1^2 - w_3^2) .
\label{04}
\end{eqnarray}
Similarly, the difference ${\rm id}(2,3)-{\rm id}(3,3)$ is used to obtain 
$G(1, 1, 2, 1, 1, 1)$,
\begin{eqnarray}
G(1, 1, 2, 1, 1, 1)\,(u_2^2 - u_3^2 - w_1^2) &=& 
                         G(1, 0, 2, 1, 1, 1) - G(1, 1, 1, 0, 2, 1) 
                 \nonumber \\ &&
                       - G(1, 1, 1, 1, 1, 1) + G(1, 1, 1, 1, 2, 0) 
                 \nonumber \\ &&
                       - G(1, 1, 2, 0, 1, 1) + 2\,G(1, 1, 1, 2, 1, 1)\,w_1^2 
                 \nonumber \\ &&
                       + G(1, 1, 1, 1, 2,1)\,(w_1^2 +w_2^2 -w_3^2) .
\label{05}
\end{eqnarray}
These two equations are used now to derive recursions in $n_2$ and $n_3$. 
With the help of the trivial formula
\begin{equation}
\int_0^\infty du\,\frac{e^{-u\,r}}{r} = \frac{1}{r^2} ,
\label{06}
\end{equation}
one integrates with respect to $u_1$, which lowers the first argument $n_1$ to $-1$.
Next, one differentiates with respect to
$u_2,u_3,w_1,w_2,w_3$ at $u_2=u_3=0$ to generate arbitrary powers 
of $r_{13}, r_{12}, r_1, r_2, r_3$ and obtains quite long recursion
relations for $n_2$ and $n_3$,
\begin{eqnarray}
f(-1,n_2+2,n_3,n_4,n_5,n_6) &=& 
    \frac{1}{w_1^2\,w_3}\,\biggl[
    n_4\,(n_4-1)\,(n_2+1)\,f(-1,n_2,n_3,n_4-2,n_5,n_6+1) 
\nonumber \\ &&
    -n_5\,(n_5-1)\,(n_2+1)\,f(-1,n_2,n_3,n_4,n_5-2,n_6+1)
\nonumber \\ &&
    +n_6\,(1+n_2+2\,n_4+n_6)\,(n_2+1)\,f(-1,n_2,n_3,n_4,n_5,n_6-1)
\nonumber \\ &&
    -n_3\,(n_3-1)\,n_6\,f(-1,n_2+2,n_3-2,n_4,n_5,n_6-1) 
\nonumber \\ &&
    +n_4\,(n_4-1)\,n_6\,f(-1,n_2+2,n_3,n_4-2,n_5,n_6-1) 
\nonumber \\ &&
   -2\,n_4\,(n_2+1)\,w_1\,f(-1,n_2,n_3,n_4-1,n_5,n_6+1) 
\nonumber \\ &&
   -2\,n_6\,(n_2+1)\,w_1\,f(-1,n_2,n_3,n_4+1,n_5,n_6-1) 
\nonumber \\ &&
   -2\,n_4\,n_6\,w_1\,f(-1,n_2+2,n_3,n_4-1,n_5,n_6-1) 
\nonumber \\ &&
   +n_6\,w_1^2\,f(-1,n_2+2,n_3,n_4,n_5,n_6-1) 
\nonumber \\ &&
   +2\,n_5\,(n_2+1)\,w_2\,f(-1,n_2,n_3,n_4,n_5-1,n_6+1) 
\nonumber \\ &&
   -(2+n_2+2\,n_4+2\,n_6)\,(n_2+1)\,w_3\,f(-1,n_2,n_3,n_4,n_5,n_6) 
\nonumber \\ &&
   +(n_3-1)\,n_3\,w_3\,f(-1,n_2+2,n_3-2,n_4,n_5,n_6)
\nonumber \\ &&
   -n_4\,(n_4-1)\,w_3\,f(-1,2+n_2,n_3,-2+n_4,n_5,n_6) 
\nonumber \\ &&
   +2\,(n_2+1)\,w_1\,w_3\,f(-1,n_2,n_3,n_4+1,n_5,n_6)\,w_1\,w_3 
\nonumber \\ &&
   +2\,n_4\,w_1\,w_3\,f(-1,n_2+2,n_3,n_4-1,n_5,n_6)
\nonumber \\ &&
   +(n_2+1)\,(w_1^2-w_2^2+w_3^2)\,f(-1,n_2,n_3,n_4,n_5,n_6+1)
\nonumber \\ &&
   +\delta_{n_4}\,w_3\,\Gamma(n_3+n_5-1,n_2+n_6+1,-1;w_2,w_3,0)
\nonumber \\ &&
   -\delta_{n_3}\,w_3\,\Gamma(-1+n_4+n_5,n_6,n_2;w_1+w_2,w_3,0)
\nonumber \\ &&
    -(n_2+n_6+1)\,\delta_{n_4}\,\Gamma(n_3+n_5-1,n_2+n_6,-1;w_2,w_3,0) 
\nonumber \\ &&
   +n_6\,\delta_{n_3}\,\Gamma(n_4+n_5-1,n_6-1,n_2;w_1+w_2,w_3,0)
\nonumber \\ &&
   +(n_2+1)\,\delta_{n_5}\,\Gamma(n_6-1,n_3+n_4-1,n_2;w_3,w_1,0)
\biggr] ,
\label{07}\\
f(-1,n_2,n_3+2,n_4,n_5,n_6) &=&\frac{1}{w_1^2\,w_2}\,\biggl[
    -n_2\,(n_2-1)\,n_5\,f(-1,-n_2-2,n_3+2,n_4,n_5-1,n_6)
\nonumber \\ &&
    +(n_3+1)\,n_4\,(n_4-1)\,f(-1,n_2,n_3,n_4-2,n_5+1,n_6) 
\nonumber \\ &&
    +(n_3+1)\,n_5\,(1+n_3+2\,n_4+n_5)\,f(-1,n_2,n_3,n_4,n_5-1,n_6) 
\nonumber \\ &&
    -(n_3+1)\,n_6\,(n_6-1)\,f(-1,n_2,n_3,n_4,n_5+1,n_6-2) 
\nonumber \\ &&
    +n_4\,(n_4-1)\,n_5\,f(-1,n_2,n_3+2,n_4-2,n_5-1,n_6) 
\nonumber \\ &&
    -2\,(n_3+1)\,n_4\,w_1\,f(-1,n_2,n_3,n_4-1,n_5+1,n_6) 
\nonumber \\ &&
    -2\,(n_3+1)n_5\,w_1\,f(-1,n_2,n_3,n_4+1,n_5-1,n_6)
\nonumber \\ &&
    -2\,n_4\,n_5\,w_1\,f(-1,n_2,2+n_3,n_4-1,n_5-1,n_6) 
\nonumber \\ &&
    +(n_3+1)\,(w_1^2+w_2^2-w_3^2)\,f(-1,n_2,n_3,n_4,n_5+1,n_6) 
\nonumber \\ &&
    +n_5\,w_1^2\,f(-1,n_2,n_3+2,n_4,n_5-1,n_6)
\nonumber \\ &&
    +n_2\,(n_2-1)\,w_2\,f(-1,n_2-2,n_3+2,n_4,n_5,n_6)
\nonumber \\ &&
    -(n_3+1)\,(n_3+2\,n_4+2\,n_5+2)\,w_2\,f(-1,n_2,n_3,n_4,n_5,n_6) 
\nonumber \\ &&
    -n_4\,(n_4-1)\,w_2\,f(-1,n_2,2+n_3,n_4-2,n_5,n_6) 
\nonumber \\ &&
    +2\,(n_3+1)\,w_1\,w_2\,f(-1,n_2,n_3,n_4+1,n_5,n_6) 
\nonumber \\ &&
    +2\,n_4\,w_1\,w_2\,f(-1,n_2,n_3+2,n_4-1,n_5,n_6) 
\nonumber \\ &&
    +2\,(n_3+1)\,n_6\,w_3\,f(-1,n_2,n_3,n_4,n_5+1,n_6-1)
\nonumber \\ &&
    +\delta_{n_4}\,w_2\,\Gamma(n_3+n_5+1,n_2+n_6-1,-1;w_2,w_3,0)
\nonumber \\ &&
    -\delta_{n_2}\,w_2\,\Gamma(n_4+n_6-1,n_5,n_3;w_1+w_3,w_2,0)
\nonumber \\ &&
    +(n_3+1)\delta_{n_6}\,\Gamma(n_2+n_4-1,n_5-1,n_3;w_1,w_2,0) 
\nonumber \\ &&
    -(1+n_3+n_5)\,\delta_{n_4}\,\Gamma(n_3+n_5,n_2+n_6-1,-1;w_2,w_3,0)
\nonumber \\ &&
    +n_5\,\delta_{n_2}\,\Gamma(n_4+n_6-1,n_5,n_3-1;w_1+w_3,w_2,0)
\biggr] ,
\label{08}
\end{eqnarray}
where $\delta_n$ denotes Kronecker delta $\delta_{n,0}$ and
$\Gamma$ is a two-electron Hylleraas integral, which is defined in Eq. (\ref{B1}).
These recursions assume that the values of $f(-1,0,0,n_4,n_5,n_6)$,
$f(-1,1,0,n_4,n_5,n_6)$, $f(-1,0,1,n_4,n_5,n_6)$ and 
$f(-1,1,1,n_4,n_5,n_6)$ are known. We calculate master integrals 
for the last three cases explicitly and express them in terms of 
two-electron Hylleraas integrals as in \cite{king_lit1},
\begin{eqnarray}
f(-1,1,1,0,0,0) &=& \int\frac{d^3\,r_1}{4\,\pi}\int\frac{d^3\,r_2}{4\,\pi}
\int\frac{d^3\,r_3}{4\,\pi}\frac{e^{-w_1\,r_1-w_2\,r_2-w_3\,r_3}}{r_{23}^2\,r_1\,r_2\,r_3}
\nonumber \\ &=& \frac{1}{w_1^2\,(w_2^2-w_3^2)}\,\ln\frac{w_2}{w_3} =
\frac{1}{w_1^2}\,\Gamma(0,0,-1;w_2,w_3,0) ,\label{09}\\ \nonumber \\
f(-1,0,1,0,0,0)  &=& \int\frac{d^3\,r_1}{4\,\pi}\int\frac{d^3\,r_2}{4\,\pi}
\int\frac{d^3\,r_3}{4\,\pi}\frac{e^{-w_1\,r_1-w_2\,r_2-w_3\,r_3}}{r_{23}^2\,r_{13}\,r_1\,r_2\,r_3}
\nonumber \\ &=& \frac{1}{4\,w_1^2\,w_2}\,
\biggl[\ln^2\biggl(\frac{w_2}{w_2 + w_3}\biggr) - \ln^2\biggl(\frac{w_2}{w_1 + w_2 + w_3}\biggr) +
  2\,{\rm Li}_2\biggl(\frac{w_2}{w_2 + w_3}\biggr) \nonumber \\ && 
 -2\,{\rm Li}_2\biggl(\frac{w_2}{w_1 + w_2 + w_3}\biggr) +
  2\,{\rm Li}_2\biggl(1 - \frac{w_3}{w_2}\biggr) - 
  2\,{\rm Li}_2\biggl(1 - \frac{w_1 + w_3}{w_2}\biggr)\biggr]\nonumber \\ &=&
\frac{1}{w_1^2}\,\Bigl[\Gamma(-1,0,-1;w_3,w_2,0)-\Gamma(-1,0,-1;w_1+w_3,w_2,0)\Bigr] ,
\label{10}\\
\nonumber \\
f(-1,1,0,0,0,0)  &=& \int\frac{d^3\,r_1}{4\,\pi}\int\frac{d^3\,r_2}{4\,\pi}
\int\frac{d^3\,r_3}{4\,\pi}\frac{e^{-w_1\,r_1-w_2\,r_2-w_3\,r_3}}{r_{23}^2\,r_{12}\,r_1\,r_2\,r_3}
\nonumber \\ &=& 
\frac{1}{w_1^2}\,\Bigl[\Gamma(-1,0,-1;w_2,w_3,0)-\Gamma(-1,0,-1;w_1+w_2,w_3,0)\Bigr] ,
\label{11}
\end{eqnarray}
where Li$_2$ is a dilogarithmic function, Eq. (\ref{C8}).
The recursion relations in $n_4, n_5, n_6$ are obtained by differentiation
with respect to $w_1, w_2$ and $w_3$
\begin{eqnarray}
f(-1,n_2,n_3,n_4,n_5,n_6) &=& \frac{1}{w_1^2}\,\biggl[
   -n_4\,(n_4-1)\,f(-1,n_2,n_3,n_4-2,n_5,n_6) 
\nonumber \\ &&
   +2\,n_4\,w_1\,f(-1,n_2,n_3,n_4-1,n_5,n_6)
\nonumber \\ &&
   +\delta_{n_4}\,\Gamma(n_5+n_3-1,n_6+n_2-1,-1;w_2,w_3,0) 
\nonumber \\ &&
   -\delta_{n_3}\,\Gamma(n_4+n_5-1,n_6,-1;w_1+w_2,w_3,0)
\nonumber \\ &&
   -\delta_{n_2}\,\Gamma(n_4+n_6-1,n_5,-1;w_1+w_3,w_2,0)\biggr] ,
\label{12}
\end{eqnarray}
for $(n_2,n_3) \in \{(1,1), (1,0),(0,1)\}$.

The calculation of the integral $f(-1,0,0,n_4,n_5,n_6)$ is much more elaborate
and we have to return to integration by parts identities, see Appendix A. These are 9
equations, which we solve against the following $X_{i=1,9}$ unknowns
at $u_2=u_3=0$,
\begin{eqnarray}
X_1 &=& G(1,2,1,1,1,1)\, u_1 , \nonumber \\
X_2 &=& G(1,1,2,1,1,1)\, u_1 , \nonumber \\
X_3 &=& G(1,1,1,1,2,1)\, u_1 , \nonumber \\
X_4 &=& G(1,1,1,1,1,2)\, u_1 , \nonumber \\
X_5 &=& G(1,2,1,1,1,1) , \nonumber \\
X_6 &=& G(1,1,2,1,1,1) , \nonumber \\
X_7 &=& G(1,1,1,1,2,1) , \nonumber \\
X_8 &=& G(1,1,1,1,1,2) , \nonumber \\
X_9 &=& G(2,1,1,1,1,1) .\label{13}
\end{eqnarray}
The solution for $X_7$ and $X_8$ is
\begin{eqnarray}
G(1, 1, 1, 1, 2, 1) &=& -G(2, 1, 1, 1, 1, 1)\,\frac{u_1^2}{2\,w^2_2} -
  G(1, 1, 1, 2, 1, 1)\,\frac{w^2_1 + w^2_2 - w^2_3}{2\,w^2_2}
\nonumber \\ &&
 +G(1, 1, 1, 1, 1, 1)\,\frac{3\,w^2_1 + w^2_2 - w^2_3}{4\,w^2_1\,w^2_2} +
  \frac{F(u_1)}{4\,w_1^2\,w_2^2} ,\label{14}
\\ \nonumber \\ 
G(1, 1, 1, 1, 1, 2) &=& -G(2, 1, 1, 1, 1, 1)\,\frac{u_1^2}{2\,w^2_3} -
  G(1, 1, 1, 2, 1, 1)\,\frac{w^2_1 - w^2_2 + w^2_3}{2\,w^2_3} 
\nonumber \\ &&
+ G(1, 1, 1, 1, 1, 1)\,\frac{3\,w^2_1 - w^2_2 + w^2_3}{4\,w^2_1\,w^2_3} -
 \frac{F(u_1)}{4\,w_1^2\,w_3^2} ,\label{15}
\end{eqnarray}
where
\begin{eqnarray}
F(u_1) &=& 2\,\biggl[G(1, 1, 1, 0, 2, 1)\,w_2^2 - G(1, 1, 1, 1, 2, 0)\,w_2^2 
\nonumber \\ &&
+ G(2, 0, 1, 1, 1, 1)\,w_2^2 - G(2, 1, 1, 1, 1, 0)\,w_2^2
\nonumber \\ &&
 - G(1, 1, 1, 0, 1, 2)\,w^2_3 + G(1, 1, 1, 1, 0, 2)\,w^2_3
\nonumber \\ &&
 - G(2, 1, 0, 1, 1, 1)\,w^2_3 + G(2, 1, 1, 1, 0, 1)\,w^2_3\biggr] .\label{16}
\end{eqnarray}
We now use explicit form of two-electron integrals in Eq. (\ref{B6})
and integrate both equations with respect to $u_1$,
\begin{eqnarray}
f(-1,0,0,0,1,0) &=& \frac{1}{2\,w_1^2\,w_2}
\Bigl[F + (2\,w_1^2 + w_2^2 - w_3^2)\,f(-1, 0, 0, 0, 0, 0) 
\nonumber \\ &&
- w_1\,(w_1^2 + w_2^2 - w_3^2)\,f(-1, 0, 0, 1, 0, 0)\Bigr] ,\label{17}\\
f(-1,0,0,0,0,1) &=& \frac{1}{2\,w_1^2\,w_3}
\Bigl[-F + (2\,w_1^2 - w_2^2 + w_3^2)\,f(-1, 0, 0, 0, 0, 0) 
\nonumber \\ &&
- w_1\,(w_1^2 - w_2^2 + w_3^2)\,f(-1, 0, 0, 1, 0, 0)\Bigr] ,\label{18}
\end{eqnarray}
where
\begin{eqnarray}
F &=& \int_0^\infty du_1\,F(u_1)  \nonumber \\ &=&\frac{1}{2}\biggl[
2\,{\rm Li}_2\biggl( -\frac{w_2}{w_1}\biggr) - {\rm Li}_2\biggl(1 - \frac{w_2}{w_3}\biggr) +
    {\rm Li}_2\biggl(1 - \frac{w_1 + w_2}{w_3}\biggr) - 2\,{\rm Li}_2\biggl(-\frac{w_3}{w_1}\biggr) 
\nonumber \\ &&
    + {\rm Li}_2\biggl(\frac{w_2}{w_2 + w_3}\biggr) - {\rm Li}_2\biggl(\frac{w_3}{w_2 + w_3}\biggr) +
    {\rm Li}_2\biggl(\frac{w_2}{w_1 + w_2 + w_3}\biggr) - 
    {\rm Li}_2\biggl(\frac{w_3}{w_1 + w_2 + w_3}\biggr) 
\nonumber \\ &&
    + {\rm Li}_2\biggl(1 - \frac{w_3}{w_2}\biggr) - {\rm Li}_2\biggl(1 -
    \frac{w_1 + w_3}{w_2}\biggr)
+\ln\biggl(\frac{w_2}{w_3}\biggr)\,\ln\biggl(\frac{w_1 + w_2 + w_3}{w_2 +
  w_3}\biggr)\biggr]\nonumber \\
&=&  \Gamma(0,-1,-2;w_2,w_1+w_3,0)- \Gamma(0,-1,-2;w_3,w_1+w_2,0) 
\nonumber \\ &&
   + \Gamma(0,-1,-2,0;w_1,w_2) - \Gamma(0,-1,-2,0;w_1,w_3)  
\nonumber \\ &&
  + w_2 \,\Gamma(-1,0,-1;w_1,0,w_2) - w_3 \,\Gamma(-1,0,-1;w_1,0,w_3) 
\nonumber \\ &&
   + w_2 \,\Gamma(-1,0,-1;0,w_2,w_3) - w_3 \,\Gamma(-1,0,-1;0,w_3,w_2)\,.\label{19}
\end{eqnarray}
Next, we  multiply both equations by powers of $w_i$
to eliminate any $w_i$ from the denominator  and differentiate with respect to
$w_1, w_2$ and $w_3$. This leads to the following recursion relations in 
$n_5$ and $n_6$ of the $f$-function
\begin{eqnarray}
f(-1,0,0,n_4,n_5+1,n_6) &=& \frac{1}{2\,w_1^2\,w_2}\Bigl[
    n_4\,(n_4-1)\,(n_4 + 2\,n_5)\,f(-1,0,0,n_4-2,n_5,n_6) 
\nonumber \\ &&
    -2\,n_4\,(n_4-1)\,w_2\,f(-1,0,0,n_4-2,n_5+1,n_6) 
\nonumber \\ &&
-n_4\,(3\,n_4 + 4\,n_5+1)\,w_1\,f(-1,0,0,n_4-1,n_5,n_6) 
\nonumber \\ &&
+4\,n_4\,w_1\,w_2\,f(-1,0,0,n_4-1,n_5+1,n_6) 
\nonumber \\ &&
+(n_4+1)\,(n_5-1)\,n_5\,f(-1,0,0,n_4,n_5-2,n_6) 
\nonumber \\ &&
- 2\,(n_4+1)\,n_5\,w_2\,f(-1,0,0,n_4,n_5-1,n_6) 
\nonumber \\ &&
- (n_4+1)\,(n_6-1)\,n_6\,f(-1,0,0,n_4,n_5,n_6-2) 
\nonumber \\ &&
+ 2\,(n_4+1)\,n_6\,w_3\,f(-1,0,0,n_4,n_5,n_6-1) 
\nonumber \\ &&
+ [(3\,n_4 + 2\,n_5+2)\,w_1^2 + (n_4+1)\,w_2^2 - (n_4+1)\,w_3^2]
\nonumber \\ &&
\times f(-1,0,0,n_4,n_5,n_6) 
\nonumber \\ &&
-(n_5-1)\,n_5\,w_1\,f(-1,0,0,n_4+1,n_5-2,n_6) 
\nonumber \\ &&
+2\,n_5\,w_1\,w_2\,f(-1,0,0,n_4+1,n_5-1,n_6) 
\nonumber \\ &&
+(n_6-1)\,n_6\,w_1\,f(-1,0,0,n_4+1,n_5,n_6-2) 
\nonumber \\ &&
-2\,n_6\,w_1\,w_3\,f(-1,0,0,n_4+1,n_5,n_6-1) 
\nonumber \\ &&
- w_1\,(w_1^2 + w_2^2 - w_3^2)\,f(-1,0,0,n_4+1,n_5,n_6) 
\nonumber \\ &&
+ F(n_4,n_5,n_6)\Bigr] ,\label{20}\\
f(-1,0,0,n_4,n_5,n_6+1) &=& \frac{1}{2\,w_1^2\,w_3}\,\Bigl[
 n_4\,(n_4-1)\,(n_4 + 2\,n_6)\,f(-1,0,0,n_4-2,n_5,n_6) 
\nonumber \\ &&
-2\,(n_4-1)\,n_4\,w_3\,f(-1,0,0,n_4-2,n_5,n_6+1) 
\nonumber \\ &&
-n_4\,(3\,n_4 + 4\,n_6 + 1)\,w_1\,f(-1,0,0,n_4-1,n_5,n_6) 
\nonumber \\ &&
+4\,n_4\,w_1\,w_3\,f(-1,0,0,n_4-1,n_5,n_6+1) 
\nonumber \\ &&
- (n_4+1)\,(n_5-1)\,n_5\,f(-1,0,0,n_4,n_5-2,n_6) 
\nonumber \\ &&
+ 2\,(n_4+1)\,n_5\,w_2\,f(-1,0,0,n_4,n_5-1,n_6) 
\nonumber \\ &&
+ (n_4+1)\,(n_6-1)\,n_6\,f(-1,0,0,n_4,n_5,n_6-2) 
\nonumber \\ &&
- 2\,(n_4+1)\,n_6\,w_3\,f(-1,0,0,n_4,n_5,n_6-1) 
\nonumber \\ &&
+ [(3\,n_4 + 2\,n_6 + 2)\,w_1^2 - (n_4+1)\,w_2^2 + (n_4+1)\,w_3^2]
\nonumber \\ &&
\times f(-1,0,0,n_4,n_5,n_6) 
\nonumber \\ &&
+(n_5-1)\,n_5\,w_1\,f(-1,0,0,n_4+1,n_5-2,n_6) 
\nonumber \\ &&
-2\,n_5\,w_1\,w_2\,f(-1,0,0,n_4+1,n_5-1,n_6) 
\nonumber \\ &&
-(n_6-1)\,n_6\,w_1\,f(-1,0,0,n_4+1,n_5,n_6-2) 
\nonumber \\ &&
+2\,n_6\,w_1\,w_3\,f(-1,0,0,n_4+1,n_5,n_6-1) 
\nonumber \\ &&
- w_1\,(w_1^2 - w_2^2 + w_3^2)\,f(-1,0,0,n_4+1,n_5,n_6) 
\nonumber \\ &&
- F(n_4,n_5,n_6)\Bigr] ,\label{21}
\end{eqnarray}
where
\begin{eqnarray}
F(n_4,n_5,n_6) &=&
(-\partial_{w_1})^{n_4}\,(-\partial_{w_2})^{n_5}\,(-\partial_{w_3})^{n_6}\,F
\nonumber \\ 
 &=&
(n_6 - n_5)\,\delta_{n_4}\,\Gamma(-1,n_5-1,n_6-1;0,w_2,w_3) 
\nonumber \\ &&
+ w_2\,\delta_{n_4}\,\Gamma(-1,n_5,n_6-1;0,w_2,w_3)
\nonumber \\ &&
-w_3\,\delta_{n_4}\,\Gamma(-1,n_6,n_5-1;0,w_3,w_2) 
\nonumber \\ &&
- (n_5-1)\,\delta_{n_6}\,\Gamma(n_4-1,0,n_5-2;w_1,0,w_2)
\nonumber \\ &&
+w_2\,\delta_{n_6}\,\Gamma(n_4-1,0,n_5-1;w_1,0,w_2) 
\nonumber \\ &&
+ (n_6-1)\,\delta_{n_5}\,\Gamma(n_4-1,0,n_6-2;w_1,0,w_3) 
\nonumber \\ &&  
-w_3\,\delta_{n_5}\,\Gamma(-1+n_4,0,n_6-1;w_1,0,w_3) 
\nonumber \\ &&
+ \Gamma(n_5,n_4+n_6-1,-2;w_2,w_1+w_3,0) 
\nonumber \\ &&  
-\Gamma(n_6,n_4+n_5-1,-2;w_3,w_1+w_2,0) .\label{22}
\end{eqnarray}
What remains is the calculation of $f(-1,0,0,n_4,0,0)$. In the following
we derive a differential equation for $h(w_1) \equiv f(-1,0,0,0,0,0)$,
from which we obtain $f(-1,0,0,n_4,0,0)$.
The solutions for 
\begin{equation}
G(2,1,1,1,1,1)=X_9 ,\label{23}
\end{equation}
and for the difference 
\begin{equation}
X_1\,u_1^{-2}-X_5=0 ,\label{24}
\end{equation}
form two algebraic equations, which however are too long to be written here.
They involve the terms $G(2, 1, 1, 1, 1, 1)$, $G(2, 1, 1, 1, 1, 1) u_1^2$,
$G(1, 1, 1, 1, 1, 1)$, $G(1, 1, 1, 1, 1, 1)\,u_1^{-2}$, $G(1, 1, 1, 2, 1, 1)$,
$G(1,1, 1, 2, 1, 1)\,u_1^{-2}$,  
and the known two-electron terms, where one of the arguments of $G$-function 
is equal to 0. We integrate both equations in $u_1$ from $\epsilon$ to
$\infty$, approach the limit $\epsilon \rightarrow 0$ and drop $\ln\epsilon$
\begin{eqnarray}
\int_\epsilon^\infty du_1\, G(2, 1, 1, 1, 1, 1) &=& -\frac{g(w_1)}{2} ,\nonumber \\
\int_\epsilon^\infty du_1\, u_1^2\,G(2, 1, 1, 1, 1, 1)  &=& \frac{h(w_1)}{2} ,\nonumber \\
\int_\epsilon^\infty du_1\, G(1, 1, 1, 1, 1, 1) &=& h(w_1) ,\nonumber \\
\int_\epsilon^\infty du_1\, u_1^{-2}\,G(1, 1, 1, 1, 1, 1) &=&
                            g(w_1)-f(1,0,0,0,0,0) , \nonumber \\
\int_\epsilon^\infty du_1\, G(1, 1, 1, 2, 1, 1) &=& -\frac{h'(w_1)}{2} ,\nonumber \\
\int_\epsilon^\infty du_1\, u_1^{-2}\, G(1,1, 1, 2, 1, 1) &=&
-\frac{g'(w_1)}{2} +\frac{1}{2}\,\frac{\partial f(1,0,0,0,0,0)}{\partial w_1} , 
\label{25}
\end{eqnarray}
where
\begin{eqnarray}
h(w_1) &=& f(-1,0,0,0,0,0) ,\nonumber \\
g(w_1) &=& \int\frac{d^3\,r_1}{4\,\pi}\int\frac{d^3\,r_2}{4\,\pi}
\int\frac{d^3\,r_3}{4\,\pi}\,e^{-w_1\,r_1-w_2\,r_2-w_3\,r_3}
(\ln r_{23}+\gamma)\,r_{31}^{-1}\,r_{12}^{-1}\,
r_{1}^{-1}\,r_{2}^{-1}\,r_{3}^{-1} ,\label{26}
\end{eqnarray}
and
\begin{equation}
f(1,0,0,0,0,0) = -\frac{1}{w_2^2\,w_3^2}\,
\ln\biggl[\frac{w_1\,(w_1+w_2+w_3)}{(w_1+w_2)\,(w_1+w_3)}\biggr] .\label{27}
\end{equation}
The set of both equations forms first order differential equations for $h(w_1)$ and
$g(w_1)$. We eliminate $g(w_1)$ to obtain the following second order differential
equation for $h(w_1)$ 
\begin{eqnarray}
&& w\,w_1^2\,h''(w_1) + 
w_1\,\bigl[4\,w_1^2\,(w_1^2 - w_2^2 - w_3^2)+w\bigr]\,h'(w_1) 
\nonumber \\
&&+\bigl[w_1^4 + 4\,w_1^2\,(w_1^2 - w_2^2 - w_3^2)-w\bigr]\,h(w_1) = R(w_1),
\label{28}
\end{eqnarray}
where
\begin{eqnarray}
R(w_1) &=& w_1\,w_2\,\ln\biggl(1 + \frac{w_1}{w_2}\biggr) + 
           w_1\,w_3\,\ln\biggl(1 + \frac{w_1}{w_3}\biggr) +
           (w_2^2 - w_3^2)\,\ln\biggl(\frac{w_1 + w_3}{w_1 + w_2}\biggr) 
\nonumber \\ &&
           +2\,w_1^2\,\ln\biggl(\frac{w_1\,(w_1 + w_2 + w_3)}{(w_1 + w_2)\,(w_1 + w_3)}\biggr) +
           (w_2^2 - w_3^2)\,F ,
\label{29}
\end{eqnarray}
and
\begin{eqnarray}
w &=& w_1^4+w_2^4+w_3^4-2\, w_1^2\, w_2^2 -2\, w_2^2\, w_3^2 -2\, w_1^2\, w_3^2\nonumber \\
  &=& -(-w_1 + w_2 + w_3) (w_1 - w_2 + w_3) (w_1 + w_2 - w_3)  (w_1 + w_2 + w_3) .
\label{30}
\end{eqnarray}
Two linearly independent solutions of the homogenous equation are:
\begin{eqnarray}
h_1(w_1) &=& \frac{1}{w_1\,\sqrt{w_2\,w_3}}\,{\rm 
  K}\biggl[\frac{(w_1+w_2-w_3)(w_1-w_2+w_3)}{4\,w_2\,w_3}\biggr] ,\nonumber \\
h_2(w_1) &=& \frac{1}{w_1\,\sqrt{w_2\,w_3}}\,{\rm
  K}\biggl[\frac{(-w_1+w_2+w_3)(w_1+w_2+w_3)}{4\,w_2\,w_3}\biggr] ,
\label{31}
\end{eqnarray}
where $K$ is a complete elliptic integral of the first kind as defined in Eq. (\ref{C1}),
and the Wronskian $W$ is
\begin{equation}
W = h_1(w_1)\,h_2'(w_1) - h_1'(w_1)\,h_2(w_1) =
\frac{2\,\pi}{w\,w_1} ,\label{32}
\end{equation}
where $w$ is defined in Eq. (\ref{30}).
The solution in Eq. (\ref{31}) is valid for $w_1$ in the range
$|w_2-w_3|<w_1<w_2+w_3$, because the elliptic integral $K$ has a branch cut
for arguments exceeding 1. We use the identity in Eq.~(\ref{D1}), 
to obtain solution $h_1$ of the homogenous equation for $w_1>w_2+w_3$
\begin{equation}
h_1(w_1) = \frac{2}{w_1\,\sqrt{(w_1+w_2-w_3)(w_1-w_2+w_3)}}\,
           K\biggl[\frac{4\,w_2\,w_3}{(w_1+w_2-w_3)(w_1-w_2+w_3)}\biggr] ,\label{35}
\end{equation}
and $h_2$ for $w_1<|w_2-w_3|$
\begin{equation}
h_2(w_1) = \frac{2}{w_1\,\sqrt{(-w_1+w_2+w_3)(w_1+w_2+w_3)}}\,
           K\biggl[\frac{4\,w_2\,w_3}{(-w_1+w_2+w_3)(w_1+w_2+w_3)}\biggr].\label{36}
\end{equation}
The solution of the inhomogeneous equation is obtained by Euler's method of 
variation of the constant,
\begin{equation}
h(w_1) = \frac{h_1(w_1)}{2\,\pi}\,\int_{w_1}^{w_2+w_3}
dw'\,\frac{R(w')\,h_2(w')}{w'} +
\frac{h_2(w_1)}{2\,\pi}\,\int_{|w_2-w_3|}^{w_1}
dw'\,\frac{R(w')\,h_1(w')}{w'} .\label{33}
\end{equation}
There is no additional term being a solution of homogenous equation,
because $h(w_1)$ is finite for all values of $w_1$, but not $h_1(w_1)$ and
$h_2(w_1)$. Therefore this is the right solution.
Having obtained $f(-1,0,0,0,0,0)\equiv h(w_1)$ and
$f(-1,0,0,1,0,0) = -h'(w_1)$ we calculate $f(-1,0,0,n_4,0,0)=h(w_1,n_4) = 
(-\partial_{w_1})^{n_4}h(w_1)$ recursively.
The inhomogeneous differential equation (\ref{28}) is differentiated
$n$-times with respect to $w_1$, to obtain
\begin{eqnarray}
h(w_1, n+2) &=& \frac{1}{w\,w_1^2}\Bigl\{-(n-3)\,(n-2)^3\,(n-1)\,n\,h(w_1,n-4) 
\nonumber \\ &&
+(n-2)\,(n-1)\,n\,(13 - 17\,n + 6\,n^2)\,w_1\,h(w_1,n-3) 
\nonumber \\ &&
-(n-1)\,n\,\Bigl[(14 - 25\,n + 15\,n^2)\,w_1^2 - 2\,(n-1)^2\,w_s\Bigr]\,h(w_1,n-2) 
\nonumber \\ &&
+2\,n\,w_1\,\Bigl[(3 - 5\,n + 10\,n^2)\,w_1^2 + (-1 + 3\,n - 4\,n^2)\,w_s\Bigr]\,h(w_1,n-1) 
\nonumber \\ &&
+ \Bigl[-(4 + 10\,n + 15\,n^2)\,w_1^4 + 
(1 - n^2)\,w_p^2 + 2\,(1 + 3\,n + 6\,n^2)\,w_1^2\,w_s\Bigr]\,h(w_1, n)
\nonumber \\ && 
+ w_1\,\Bigl[(5 + 6\,n)\,w_1^4 + (1 + 2\,n)\,w_p^2 - 2\,(3 + 4\,n)\,w_1^2\,w_s\Bigr]\,
  h(w_1,n+1) 
\nonumber \\ && 
+ R(w_1,n)\Bigr\} ,\label{39}
\end{eqnarray}
where $w_s = w_2^2+w_3^2$, $w_p = w_2^2-w_3^2$ and
\begin{eqnarray}
R(w_1,n) &=& (-\partial_{w_1})^n\,R(w_1)\nonumber \\ 
         &=& -\frac{4\,(n-3)!}{w_1^{n-2}}+\frac{5\,w_2\,(n-2)!}{(w_1+w_2)^{n-1}}
             +\frac{4\,(n-3)!}{(w_1+w_2)^{n-2}}+
             \frac{5\,w_3\,(n-2)!}{(w_1+w_3)^{n-1}}
\nonumber \\ &&
             +\frac{4\,(n-3)!}{(w_1+w_3)^{n-2}} 
             - \frac{2\,(w_2+w_3)^2\,(n-1)!}{(w_1+w_2+w_3)^n} 
             - \frac{4\,(w_2+w_3)\,(n-2)!}{(w_1+w_2+w_3)^{n-1}} 
\nonumber \\ &&
             -\frac{4\,(n-3)!}{(w_1+w_2+w_3)^{n-2}}
             +\frac{(4\,w_2^2-w_3^2)\,(n-1)!}{(w_1+w_2)^n}
             +\frac{(4\,w_3^2-w_2^2)\,(n-1)!}{(w_1+w_3)^n}
\nonumber \\ &&
             +(w_2^2-w_3^2)\,F(n,0,0),\;\;\;\mbox{\rm for}\; n>2,
\nonumber \\ 
R(w_1,0) &=& R(w_1) , \nonumber \\ 
R(w_1,1) &=& -(w_2 + w_3)+\frac{4\,w_2^2 - w_3^2}{w_1+w_2} - 2\,\frac{(w_2+w_3)^2)}{w_1+w_2+w_3} 
            +\frac{4\,w_3^2-w_2^2}{w_1+w_3} - w_2\,\ln\biggl(1+\frac{w_1}{w_2}\biggr) 
\nonumber \\ &&
            - w_3\,\ln\biggl(1+\frac{w_1}{w_3}\biggr) -
             4\,w_1\,\ln\biggl[\frac{w_1\,(w_1+w_2+w_3)}{(w_1+w_2)\,(w_1+w_3)}\biggr]
            +(w_2^2-w_3^2)\,F(1,0,0) ,
\nonumber \\ 
R(w_1,2) &=& \frac{5\,w_2}{w_1+w_2}+\frac{5\,w_3}{w_1+w_3} - \frac{2\,(w_2+w_3)^2}{(w_1+w_2+w_3)^2} 
             - \frac{4\,(w_2+w_3)}{w_1+w_2+w_3}
\nonumber \\ &&
             +\frac{4\,w_2^2-w_3^2}{(w_1+w_2)^2} +\frac{4\,w_3^2-w_2^2}{(w_1+w_3)^2}
             +4\,\ln\biggl[\frac{w_1\,(w_1+w_2+w_3)}{(w_1+w_2)\,(w_1+w_3)}\biggr]
\nonumber \\ &&
             +(w_2^2-w_3^2)\,F(2,0,0) .\label{40}
\end{eqnarray}
In the case $w_1\approx w_{\rm sing} = w_2+w_3$ or $|w_2-w_3|$,
the recursion in Eq. (\ref{39}) is not numerically
stable. Therefore, one instead of this recursion,
calculates the recursion exactly at $w_1=w_{\rm sing}$, 
what corresponds to setting $w=0$ in Eq. (\ref{39}), 
\begin{eqnarray}
h(w_1,n+1) &=& 
\frac{-1}{w_1\,\Bigl[(5 + 6\,n)\,w_1^4 + (1 + 2\,n)\,w_p^2 - 2\,(3 + 4\,n)\,w_1^2\,w_s\Bigr]}
\nonumber \\ &&
\Bigl\{-(n-3)\,(n-2)^3\,(n-1)\,n\,h(w_1,n-4) 
\nonumber \\ &&
+(n-2)\,(n-1)\,n\,(13 - 17\,n + 6\,n^2)\,w_1\,h(w_1,n-3) 
\nonumber \\ &&
-(n-1)\,n\,\Bigl[(14 - 25\,n + 15\,n^2)\,w_1^2 - 2\,(n-1)^2\,w_s\Bigr]\,h(w_1,n-2) 
\nonumber \\ &&
+2\,n\,w_1\,\Bigl[(3 - 5\,n + 10\,n^2)\,w_1^2 + (-1 + 3\,n - 4\,n^2)\,w_s\Bigr]\,h(w_1,n-1) 
\nonumber \\ &&
+ \Bigl[-(4 + 10\,n + 15\,n^2)\,w_1^4 + 
(1 - n^2)\,w_p^2 + 2\,(1 + 3\,n + 6\,n^2)\,w_1^2\,w_s\Bigr]\,h(w_1, n)
\nonumber \\ && 
+ R(w_1,n)\Bigr\} ,\label{41}
\end{eqnarray}
where $w_1=w_{\rm sing}$. This completes the recursion relations 
for the extended Hylleraas integral with $r_{23}^{-2}$.

\section{Numerical evaluation}
We pass now to numerical implementation of recursions
and integration of the master integral in Eq. (\ref{33}).
All the computation is performed with extended precision arithmetic,
namely quad and sextuple precision. Even higher precision, the octuple one, 
is used for checking numerical accuracy. The starting point is the master
integral. One needs to calculate it with the highest possible accuracy,
because the recursions depend most significantly on the value of initial terms.
The integrand in Eq. (\ref{33})
is a product of the function $R$ and the complete elliptic integral $K$.
The function $R$ defined in Eq. (\ref{29}) has singularities only at $w_i=0$ and $w_i=\infty$, 
and the complete elliptic integral has logarithmic
singularities at $w_1=w_{\rm sing} \equiv |w_2-w_3|$ or $w_2+w_3$ which 
correspond to zeros of $w$ in Eq. (\ref{30}).
In the following we assume that $w_2-w_3\neq 0$.
When $|w_2-w_3|+\epsilon_1<w_1<w_2+w_3-\epsilon_2$, the integral in
Eq. (\ref{33}) can be performed by the Gauss-Legendre quadrature \cite{nr}. 
We have verified that for $\epsilon_i\approx 0.2$ the integration
with 100 points gives the accuracy of at least 32 digits if not more.
For the cases $w_1>w_2+w_3+\epsilon_2$ and $w_1<|w_2-w_3|-\epsilon_1$,  the integration contour
is deformed on the complex plane to avoid the singularity.
So this contour consists of two lines on a real axis, joined by a half-circle
with origin at the singular point and integration  
is performed independently on each part using the Gauss-Legendre quadrature.
When $w_1$ is close to the singular point, we first obtain 
$h(w_{\rm sing}\pm\epsilon)$, next we calculate $h(w_{\rm sing})$ by matching
the Taylor expansion from recursion in Eq. (\ref{41}) with 
$h(w_{\rm sing}\pm\epsilon)$ and in the last step we again use
recursion in Eq. (\ref{41}) to obtain $h(w_1)$.
For $w_2-w_3 = 0$ we separate $R$ in Eq. (\ref{29}) into 
the part which is free of $\ln w_1$ and the part which is proportional to 
$\ln w_1$. The first part is integrated using the Gauss-Legendre quadrature
an the second one is integrated using the Gauss quadrature adapted for 
the  logarithmic weight function. Several numerical results for some selected
$w_i$  are presented in Table I.
\begin{table}[!hbt]
\caption{Values of the master integral for selected $w_1,w_2$, and $w_3$}
\label{table1}
\begin{ruledtabular}
\begin{tabular}{llll}
      $w_1$ & $w_2$  & $w_3$  &  $f(-1,0,0,0,0,0)$  \\
\hline
4.0 & 1.0 & 0.5 &   1.243\,735\,828\,073\,620\,173\,310\,981\,564\,244[-1] \\
4.0 & 1.0 & 1.0 &   9.855\,133\,136\,060\,504\,470\,218\,647\,797\,889[-2] \\
4.0 & 1.0 & 1.5 &   8.181\,412\,007\,841\,597\,436\,460\,514\,476\,518[-2] \\
4.0 & 1.0 & 2.0 &   6.983\,588\,391\,604\,680\,181\,982\,031\,823\,035[-2] \\
4.0 & 1.0 & 2.5 &   6.077\,218\,287\,692\,100\,226\,048\,417\,176\,715[-2] \\
4.0 & 1.0 & 3.0 &   5.365\,400\,720\,042\,544\,709\,716\,176\,264\,176[-2] \\
4.0 & 1.0 & 3.5 &   4.791\,010\,346\,652\,078\,406\,517\,300\,908\,585[-2] \\
4.0 & 1.0 & 4.0 &   4.317\,729\,831\,064\,450\,749\,511\,048\,756\,748[-2] \\
4.0 & 1.0 & 4.5 &   3.921\,185\,585\,221\,614\,693\,378\,573\,393\,156[-2] \\
4.0 & 1.0 & 5.0 &   3.584\,332\,630\,993\,527\,980\,351\,431\,968\,712[-2] \\
4.0 & 1.0 & 5.5 &   3.294\,856\,745\,699\,432\,037\,984\,459\,599\,008[-2] \\
\end{tabular}
\end{ruledtabular}
\end{table}

Considering recursions, all but one involve denominators limited from below.
Only that in Eq. (\ref{39}) for increasing $n_4$ has a
denominator which can be arbitrarily close to zero. Therefore, 
if $w_1\approx w_{\rm sing}$, one instead of recursion in Eq. (\ref{39}),
uses the  recursion in Eq. (\ref{41})
and calculates $h(w_1,n)$ from Taylor expansion at $w_1=w_{\rm sing}$.
All other recursions are calculated directly as in
Eqs. (\ref{07}, \ref{08}, \ref{12}, \ref{20}, \ref{21}). They involve
two-electron integrals $\Gamma$. The calculation of $\Gamma$ including 
singular cases has recently been described in detail in Refs. \cite{kor1} and \cite{harris}, 
and it does not pose any problem.  Finally, several numerical results for 
three-electron integral involving powers of $r_i$ and $r_{ij}$ 
are presented in Table II. For comparison with the former results 
obtained in Ref. \cite{king_lit3}, we have taken the same $n_i$ and $w_i$
as in Table II of this Ref. Our results agree
up to the precision achieved in  Ref. \cite{king_lit3}
with the one correction. In the fifth position instead of
$I(2,1,1,3,3,-2,4.338,4.338,7.384)$, it should be
$I(1,1,2,3,3,-2,4.338,4.338,7.384)$.
\begin{table}[!hbt]
\caption{Three-electron Hylleraas integral involving $1/r_{12}^2$.
         $n_i$ and $w_i$ are from Table II of Ref. \cite{king_lit3}.
         Function $I$ from this Ref. should be divided by $(4\,\pi)^3$ 
         for comparison with our function $f$.}
\label{table2}
\begin{ruledtabular}
\begin{tabular}{rrrrl}
                           &$w_1$ &$w_2$ &$w_3$ &  \\ \hline
         $f(-1,2,2,2,2,2)$ &2.700 &2.700 &2.700 &    3.622\,072\,193\,238\,069\,065\,841\,911\,460\,566[-3] \\
         $f(-1,2,4,1,1,1)$ &2.700 &2.900 &0.650 &    2.044\,941\,897\,990\,188\,175\,637\,070\,889\,313[-1] \\
         $f(-1,0,2,1,1,1)$ &2.700 &2.900 &0.650 &    8.560\,152\,684\,198\,427\,372\,519\,849\,562\,718[-3] \\
         $f(-1,0,0,1,1,1)$ &2.700 &2.900 &0.650 &    7.695\,548\,443\,927\,856\,456\,193\,296\,733\,495[-3] \\
         $f(-1,4,4,3,2,2)$ &7.384 &4.338 &4.338 &    2.516\,457\,130\,304\,929\,175\,434\,829\,560\,592[-6] \\
         $f(-1,0,0,2,2,2)$ &3.000 &2.000 &1.000 &    7.759\,319\,533\,814\,226\,728\,190\,558\,692\,235[-3] \\
         $f(-1,0,0,1,3,2)$ &3.000 &1.000 &2.000 &    1.528\,428\,874\,506\,937\,507\,531\,543\,743\,291[-2] \\
         $f(-1,0,4,2,4,3)$ &2.000 &3.000 &4.000 &    6.208\,037\,315\,282\,433\,323\,108\,011\,184\,899[-3] \\
         $f(-1,2,2,1,1,1)$ &2.700 &2.900 &0.650 &    4.036\,629\,272\,285\,446\,411\,970\,138\,933\,470[-2] \\
         $f(-1,2,2,1,1,1)$ &2.500 &2.500 &0.600 &    1.025\,702\,855\,657\,754\,018\,359\,340\,659\,240[-1] \\
         $f(-1,2,2,0,0,0)$ &2.700 &2.900 &0.650 &    3.674\,068\,373\,009\,625\,515\,617\,159\,197\,784[-2] \\
         $f(-1,2,2,1,2,3)$ &1.000 &1.000 &1.000 &    4.576\,295\,463\,451\,984\,514\,935\,097\,879\,411[2] \\
         $f(-1,2,2,2,2,2)$ &1.000 &1.000 &1.000 &    5.436\,536\,048\,634\,697\,021\,325\,813\,246\,683[2] \\
         $f(-1,2,4,4,3,1)$ &3.000 &2.000 &1.000 &    6.126\,463\,692\,215\,932\,446\,059\,888\,955\,061[0] \\
         $f(-1,2,4,1,0,0)$ &1.000 &1.000 &1.000 &    4.219\,398\,540\,932\,754\,336\,898\,663\,066\,822[2] \\
         $f(-1,4,4,1,1,1)$ &2.700 &2.900 &0.650 &    3.826\,213\,635\,276\,544\,192\,395\,399\,453\,200[0] \\
         $f(-1,4,6,1,1,1)$ &2.700 &2.900 &0.650 &    4.206\,326\,264\,336\,604\,338\,380\,540\,655\,410[1] \\
         $f(-1,6,6,1,1,1)$ &2.700 &2.900 &0.650 &    1.886\,948\,258\,407\,236\,970\,462\,772\,961\,418[3] 
\end{tabular}
\end{ruledtabular}
\end{table}

Considering the extended Hylleraas integral with $r_1^{-2}$, we calculate it
by numerical integration with respect to $w_1$
\begin{equation}
f(n_1,n_2,n_3,-1,n_5,n_6) = \int_{w_1}^\infty dw_1\,f(n_1,n_2,n_3,0,n_5,n_6) .
\label{42}
\end{equation}
The recursion relations for $f(n_1,n_2,n_3,n_4,n_5,n_6)$ with nonnegative
$n_i$ have been derived previously \cite{rec}, and they seem to be stable
enough to perform this integration
numerically. Since, we have not found in the literature the method of
integration, which is adapted to two weight functions: 
the constant and the logarithmic one, we use the standard Gauss-Legendre quadrature.
In Table III we present several numerical results. It is observed the significant loss 
of precision due to the presence of $\ln w_1/w_1^n$ at large $w_1$ asymptotic.
Therefore, precise integration requires in some cases, the subtraction
of these terms.
\begin{table}[!hbt]
\caption{Three-electron Hylleraas integral involving $1/r_{1}^2$ at $w_1=2, w_2=3, w_3=4$}
\label{table3}
\begin{ruledtabular}
\begin{tabular}{rl}
     $f(0,0,0,-1,0,0)$ &      5.112\,034\,507\,187\,907\,543[-2] \\
     $f(0,1,0,-1,0,0)$ &      1.376\,985\,263\,507\,039\,164[-2] \\
     $f(0,2,0,-1,0,0)$ &      6.942\,269\,369\,095\,712\,105[-3] \\
     $f(0,3,0,-1,0,0)$ &      5.403\,223\,451\,815\,895\,118[-3] \\
     $f(0,4,0,-1,0,0)$ &      5.907\,661\,306\,554\,417\,555[-3] \\
     $f(0,5,0,-1,0,0)$ &      8.587\,459\,945\,883\,427\,557[-3] \\
     $f(0,6,0,-1,0,0)$ &      1.598\,496\,287\,482\,975\,980[-2] \\
     $f(0,7,0,-1,0,0)$ &      3.698\,745\,219\,132\,481\,190[-2] \\
     $f(0,8,0,-1,0,0)$ &      1.036\,442\,843\,454\,920\,448[-1] \\
     $f(0,9,0,-1,0,0)$ &      3.434\,721\,508\,609\,856\,189[-1]
\end{tabular}
\end{ruledtabular}
\end{table}

\section{Summary}
An analytic approach is presented for the calculation of three-electron
Hylleraas integrals involving one inverse quadratic power of inter-particle
coordinate. This approach is  based on exact recursion relations in powers
of coordinates. These recursions involve initial terms and
two-electron integrals. For the initial term $f(-1,0,0,0,0,0)$
as a function of $w_1$, one constructs a linear second order differential 
equation. Its solution is expressed as a one-dimensional integral
over dilogarithmic and elliptic function $K$, which can be obtained
numerically with arbitrary high precision.
The two-electron Hylleraas integrals have already been derived in the literature
and they also can be obtained with arbitrary high precision.

These extended Hylleraas integrals are necessary for the calculation
of relativistic effects in lithium and light lithium-like ions 
\cite{king_rev,hyl_lit, lit_iso, lit_g}.
One interesting goal is the high precision calculation of the lithium hyperfine
splitting \cite{lit_hfs}, which can serve as a benchmark result 
for other less accurate methods.
Moreover, it has recently become possible to derive nuclear parameters such as
the charge radius from the measurement of the isotope shift \cite{lit_iso}. The hyperfine
splitting is sensitive to the distribution of magnetic moment.
Therefore the measurement of hfs in various odd isotopes of Li or light lithium-like
ions may lead to the determination of the so called magnetic radius, which is very difficult
to access experimentally.
Even more interesting is the possible extension of this analytic method to
beryllium and beryllium-like ions, the 4-electron systems. 
The use of Hylleraas basis set will allow for 
a high precision calculation of the wave function and, for example, various
transition rates which are of astrophysical relevance.

\section{Acknowledgments}
We are grateful to Vladimir Korobov for his source code
of the fast multi-precision arithmetics and the dilogarithmic function. 
We wish to thank Krzysztof Meissner for help in solving differential equation
and Frederick King for presenting us his numerical results for
some selected values of extended Hylleraas integral and for useful comments.
This work was supported by EU grant HPRI-CT-2001-50034.

\appendix
\renewcommand{\theequation}{\Alph{section}\arabic{equation}}
\renewcommand{\thesection}{\Alph{section}}

\section{Integration by parts identities}
The complete set of recursion relations as obtained from 
integration by parts identities is presented below.
Function $G$ is defined in Eq. (\ref{02}) and id$(i,j)= 0$ for $i,j=1,2,3$, 
\begin{eqnarray}
{\rm id}(1,1)&=&
-G(0,1,1,1,1,2) - G(0,1,1,1,2,1) + G(1,0,1,1,1,2) + G(1,1,0,1,2,1) 
\nonumber \\ &&
- G(1,1,1,1,1,1) + 2\,G(2,1,1,1,1,1)\,u_1^2 +G(1,1,1,1,2,1)\,(u_1^2 - u_3^2 + w_2^2) 
\nonumber \\ &&
+ G(1,1,1,1,1,2)\,(u_1^2 - u_2^2 + w_3^2)  ,
\nonumber \\ 
{\rm id}(2,1)&=&
-G(0,1,1,1,1,2) - G(0,1,1,1,2,1) + G(1,0,1,1,1,2) + G(1,1,0,1,2,1) 
\nonumber \\ &&
- G(1,1,1,0,2,1) + G(1,1,1,1,2,0) - G(2,0,1,1,1,1) + G(2,1,1,1,1,0) 
\nonumber \\ &&
+ G(1,1,1,1,1,2)\,(u_1^2 - u_2^2 - w_3^2) +
  G(2,1,1,1,1,1)\,(u_1^2 + u_2^2 - w_3^2) 
\nonumber \\ &&
+ G(1,1,1,1,2,1)\,(u_1^2 - u_3^2+ w_1^2 - w_3^2) ,
\nonumber \\ 
{\rm id}(3,1)&=&
-G(0,1,1,1,1,2) -G(0,1,1,1,2,1) + G(1,0,1,1,1,2) + G(1,1,0,1,2,1) 
\nonumber \\ &&
- G(1,1,1,0,1,2) + G(1,1,1,1,0,2) - G(2,1,0,1,1,1) + G(2,1,1,1,0,1) 
\nonumber \\ &&
+ G(1,1,1,1,2,1)\,(u_1^2 - u_3^2 - w_2^2) + G(2,1,1,1,1,1)\,(u_1^2 + u_3^2 - w_2^2) 
\nonumber \\ &&
+ G(1,1,1,1,1,2)\,(u_1^2 - u_2^2 + w_1^2 - w_2^2) ,
\nonumber \\  
{\rm id}(2,2)&=&
G(0,1,1,1,1,2) - G(1,0,1,1,1,2) - G(1,0,1,2,1,1) + G(1,1,0,2,1,1) 
\nonumber \\ &&
- G(1,1,1,1,1,1) + 2\,G(1,2,1,1,1,1)\,u_2^2 + G(1,1,1,2,1,1)\,(u_2^2 - u_3^2 + w_1^2) 
\nonumber \\ &&
+ G(1,1,1,1,1,2)\,(-u_1^2 + u_2^2 + w_3^2) ,
\nonumber \\ 
{\rm id}(1,2)&=&
G(0,1,1,1,1,2) - G(0,2,1,1,1,1) - G(1,0,1,1,1,2) - G(1,0,1,2,1,1) 
\nonumber \\ &&
+ G(1,1,0,2,1,1) - G(1,1,1,2,0,1) + G(1,1,1,2,1,0) + G(1,2,1,1,1,0) 
\nonumber \\ &&
+ G(1,1,1,1,1,2)\,(-u_1^2 + u_2^2 - w_3^2) + G(1,2,1,1,1,1)\,(u_1^2 + u_2^2 -w_3^2) 
\nonumber \\ &&
+ G(1,1,1,2,1,1)\,(u_2^2 - u_3^2 + w_2^2 - w_3^2)  ,
\nonumber \\ 
{\rm id}(3,2)&=&
G(0,1,1,1,1,2) - G(1,0,1,1,1,2) - G(1,0,1,2,1,1) + G(1,1,0,2,1,1) 
\nonumber \\ &&
+G(1,1,1,0,1,2) - G(1,1,1,1,0,2) - G(1,2,0,1,1,1) + G(1,2,1,0,1,1) 
\nonumber \\ &&
+ G(1,1,1,2,1,1)\,(u_2^2 - u_3^2 - w_1^2) + G(1,2,1,1,1,1)\,(u_2^2 + u_3^2 - w_1^2) 
\nonumber \\ &&
+ G(1,1,1,1,1,2)\,(-u_1^2 + u_2^2 - w_1^2 + w_2^2) ,
\nonumber \\ 
{\rm id}(3,3)&=&
G(0,1,1,1,2,1) + G(1,0,1,2,1,1) - G(1,1,0,1,2,1) - G(1,1,0,2,1,1) 
\nonumber \\ &&
- G(1,1,1,1,1,1) + 2\,G(1,1,2,1,1,1)\,u_3^2 + G(1,1,1,2,1,1)\,(-u_2^2 + u_3^2 + w_1^2) 
\nonumber \\ &&
+ G(1,1,1,1,2,1)\,(-u_1^2 + u_3^2+ w_2^2) ,
\nonumber \\  
{\rm id}(2,3)&=&
G(0,1,1,1,2,1) + G(1,0,1,2,1,1) - G(1,0,2,1,1,1) - G(1,1,0,1,2,1) 
\nonumber \\ &&
- G(1,1,0,2,1,1) + G(1,1,1,0,2,1) - G(1,1,1,1,2,0) + G(1,1,2,0,1,1) 
\nonumber \\ &&
+ G(1,1,1,2,1,1)\,(-u_2^2 + u_3^2 - w_1^2) + G(1,1,2,1,1,1)\,(u_2^2 + u_3^2 - w_1^2) 
\nonumber \\ &&
+ G(1,1,1,1,2,1)\,(-u_1^2 + u_3^2- w_1^2 + w_3^2) , \nonumber \\ 
{\rm id}(1,3)&=&
G(0,1,1,1,2,1) - G(0,1,2,1,1,1) + G(1,0,1,2,1,1) - G(1,1,0,1,2,1) 
\nonumber \\ &&
- G(1,1,0,2,1,1) + G(1,1,1,2,0,1) - G(1,1,1,2,1,0) + G(1,1,2,1,0,1) 
\nonumber \\ &&
+ G(1,1,1,1,2,1)\,(-u_1^2 + u_3^2 - w_2^2) + G(1,1,2,1,1,1)\,(u_1^2 + u_3^2 - w_2^2) 
\nonumber \\ &&
+ G(1,1,1,2,1,1)\,(-u_2^2 + u_3^2 - w_2^2 + w_3^2) . \nonumber \\ 
\label{A1}
\end{eqnarray}

\section{Two-electron integrals}
\setcounter{equation}{0}
\noindent The two-electron integral $\Gamma$ is defined by
\begin{equation}
\Gamma(n_1,n_2,n_3;\alpha_1,\alpha_2,\alpha_3) =  \int\frac{d^3\,r_1}{4\,\pi}
\int\frac{d^3\,r_2}{4\,\pi}\,e^{-\alpha_1\,r_1-\alpha_2\,r_2-\alpha_3\,r_{12}}\,
r_{1}^{n_1-1}\,r_{2}^{n_2-1}\,r_{12}^{n_3-1} .\label{B1}
\end{equation}
In the simplest case of $n_1=n_2=n_3=0$ it is 
\begin{equation}
\Gamma(0,0,0,;\alpha_1,\alpha_2,\alpha_3) =
\frac{1}{(\alpha_1+\alpha_2)(\alpha_2+\alpha_3)(\alpha_3+\alpha_1)} .\label{B2}
\end{equation}
In the general case of $n_i\geq0$
\begin{equation}
\Gamma(n_1,n_2,n_3;\alpha_1,\alpha_2,\alpha_3) = 
\biggl(-\frac{\rm d}{{\rm d}\alpha_1}\biggr)^{n_1}\,
\biggl(-\frac{\rm d}{{\rm d}\alpha_2}\biggr)^{n_2}\,
\biggl(-\frac{\rm d}{{\rm d}\alpha_3}\biggr)^{n_3}\,
\frac{1}{(\alpha_1+\alpha_2)(\alpha_2+\alpha_3)(\alpha_3+\alpha_1)},\label{B3}
\end{equation}
and recursions relations for its evaluations have been derived in Refs. 
\cite{kolos}.
The two-electron integral $\Gamma$ for $n_i<0$ can be obtained by the
integration over $\alpha_i$. Typical examples are
\begin{eqnarray}
\Gamma(-1,n_2,n_3;\alpha_1,\alpha_2,\alpha_3) &=& 
\biggl(-\frac{\rm d}{{\rm d}\alpha_2}\biggr)^{n_2}\,
\biggl(-\frac{\rm d}{{\rm d}\alpha_3}\biggr)^{n_3}\,
\frac{\ln(\alpha_1+\alpha_2)-\ln(\alpha_1+\alpha_3)}
{(\alpha_2-\alpha_3)(\alpha_2+\alpha_3)} , \label{B4}\\
\Gamma(-1,n_2,-1;\alpha_1,\alpha_2,\alpha_3) &=&
\biggl(-\frac{\rm d}{{\rm d}\alpha_2}\biggr)^{n_2}\,
\frac{1}{2\,\alpha_2}\,\bigg[\frac{\pi^2}{6}+\frac{1}{2}\,
\ln^2\biggl(\frac{\alpha_1+\alpha_2}{\alpha_2+\alpha_3}\biggr)
\nonumber \\ &&
+{\rm Li}_2\biggl(1-\frac{\alpha_1+\alpha_3}{\alpha_1+\alpha_2}\biggr)
+{\rm Li}_2\biggl(1-\frac{\alpha_1+\alpha_3}{\alpha_2+\alpha_3}\biggr)
\biggr] .\label{B5}
\end{eqnarray}
Other examples together with recursion relations to calculate derivatives
may be found in Refs. \cite{kor1, harris}.
Some three-electron integrals $G$ can be expressed in terms of $\Gamma$.
It is when any of argument is equal to zero. Complete list of all cases is:
\begin{eqnarray}
G(0,1,1,1,1,1) &=& \Gamma(-1,0,-1;w_2+w_3,w_1,u_2+u_3) , \nonumber\\
G(1,0,1,1,1,1) &=& \Gamma(-1,0,-1;w_1+w_3,w_2,u_1+u_3) , \nonumber\\
G(1,1,0,1,1,1) &=& \Gamma(-1,0,-1;w_1+w_2,w_3,u_1+u_2) , \nonumber\\
G(1,1,1,0,1,1) &=& \Gamma(-1,0,-1;w_2+u_3,u_1,w_3+u_2) , \nonumber\\
G(1,1,1,1,0,1) &=& \Gamma(-1,0,-1;w_1+u_3,u_2;w_3+u_1) , \nonumber\\
G(1,1,1,1,1,0) &=& \Gamma(-1,0,-1;w_1+u_2,u_3,w_2+u_1) . \label{B6}
\end{eqnarray}

\section{Special functions}
\setcounter{equation}{0}
The complete elliptic integral of the first and second kind, $K$ and $E$
respectively, are defined according to \cite{abram} as
\begin{eqnarray}
K(m) &=& \int_0^1 dt\,(1-t^2)^{-1/2}\,(1-m\,t^2)^{-1/2} ,\label{C1}\\
E(m) &=& \int_0^1 dt\,(1-t^2)^{-1/2}\,(1-m\,t^2)^{1/2} .\label{C2}
\end{eqnarray}
They are related to a hypergeometric function
\begin{eqnarray}
K(m) &=& \frac{\pi}{2}\,_2F_1(1/2,1/2;1;m) ,\label{C3}\\
E(m) &=&  \frac{\pi}{2}\,_2F_1(-1/2,1/2;1;m) ,\label{C4}
\end{eqnarray}
and fulfill the Legendre's relation
\begin{equation}
E(m)\,K(1-m)+E(1-m)\,K(m)-K(m)\,K(1-m) = \frac{\pi}{2} .\label{C5}
\end{equation} 
Their first derivatives are
\begin{eqnarray}
K'(m) &=& \frac{E(m)}{2\,m\,(1-m)} - \frac{K(m)}{2\,m} ,\label{C6}\\
E'(m) &=& \frac{E(m)-K(m)}{2\,m} .\label{C7}
\end{eqnarray}
Elliptic functions for $|m| \le 1 $ can be conveniently calculated numerically
as described in \cite{nr}. For $m>1$ one uses the identity \cite{abram}
\begin{eqnarray}
K(m\pm i\,\epsilon) &=&  \frac{1}{\sqrt{m}}\,K\biggl(\frac{1}{m}\biggr) 
\pm i\,K\bigl(1-m\bigr) , \label{D1} \\
E(m\pm i\,\epsilon) &=&
\sqrt{m}\,E\biggl(\frac{1}{m}\biggr)+\frac{1-m}{\sqrt{m}}\,K\biggl(\frac{1}{m}\biggr)
\pm i\,\bigl[K(1-m)-E(1-m)\bigr] ,
\end{eqnarray}
and for $m<-1$ 
\begin{eqnarray}
K(m) &=&\frac{1}{\sqrt{1-m}}\,K\biggl(\frac{m}{m-1}\biggr) \label{D2}\\
E(m) &=&{\sqrt{1-m}}\,E\biggl(\frac{m}{m-1}\biggr) . 
\end{eqnarray}
The Laurent expansion near the singularity $m=1$ is
\begin{equation}
K(m)  =
\sum_{n=0}^\infty\,\biggl[\frac{1}{n!}\,\biggl(\frac{1}{2}\biggr)_n\biggr]^2\,
\biggl[\psi(n+1) - \psi(n+1/2)-
\frac{1}{2}\,\ln(1-m)\biggr]\,(1-m)^n , \label{expansion}
\end{equation}
where $\psi$ is a logarithmic derivative of the Euler Gamma function.

The dilogarithmic function Li$_2$ is defined by
\begin{equation}
{\rm Li}_2(z) = -\int_0^z \frac{\ln(1-z)}{z} .\label{C8}
\end{equation}
Taylor expansion around origin 
\begin{equation}
{\rm Li}_2(z) = \sum_{i=1}^\infty \frac{z^i}{i^2} , \label{C9}
\end{equation}
is convergent for $|z| \leq 1$.
Two useful relations
\begin{eqnarray}
{\rm Li}_2(-x) + {\rm Li}_2\bigg(-\frac{1}{x}\biggr) &=& 
-\frac{\pi^2}{6}-\frac{\ln^2 x}{2} ,\label{C10}\\
{\rm Li}_2(x) + {\rm Li}_2(1-x) &=& 
\frac{\pi^2}{6}-\ln x\,\ln(1-x) ,\label{C11}
\end{eqnarray} 
are used for simplification of result of integrations in Eqs.(\ref{10}, \ref{19}).
Further formulas may be found in \cite{abram} and an efficient numerical evaluation
is described among others in \cite{kor1}.
\end{document}